\documentclass[12pt]{article}
\pdfoutput=1
\usepackage{amsmath}
\usepackage{longtable}
\usepackage{graphicx}

\newcommand{\be}{\begin{equation}}
\newcommand{\ee}{\end{equation}}

\newcommand{\bean}{\begin{eqnarray}}
\newcommand{\eean}{\end{eqnarray}}

\newcommand{\bea}{\begin{eqnarray*}}
\newcommand{\eea}{\end{eqnarray*}}

\def\Sum{\sum\limits}

\def\art#1{[\ref{#1}]}

\begin{document}

\begin{center}

{\bf Detecting somatic mutations in genomic sequences by means of Kolmogorov-Arnold analysis}

\vspace{0.2in}

\noindent{V. G.~Gurzadyan$^{1,2}$, H. Yan$^3$, G. Vlahovic$^3$, \\ A. Kashin$^{2}$, P. Killela$^{3}$, Z. Reitman$^{3}$,\\
S. Sargsyan$^{2}$, G. Yegorian$^{2}$, G.  Milledge$^{1}$, B. Vlahovic$^1$
}
\end{center}

\vspace{0.2in}

{\footnotesize\it
\noindent
$^1$  NSF Computational Center of Research Excellence and NASA University Research Center for Aerospace Device, NCCU, Durham, NC, USA\\
$^2$ Yerevan Physics Institute and Yerevan State University, Yerevan, Armenia\\
$^3$ Department of Pathology, Duke University Medical Center, The Preston Robert Tisch Brain Tumor Center at Duke, and
    Pediatric Brain Tumor Foundation Institute at Duke, Durham, NC, USA
}


\vspace{0.2in}

{\bf Abstract}
The Kolmogorov-Arnold stochasticity parameter technique is applied for the first time to the study of cancer genome sequencing, to reveal mutations. Using data generated by next generation sequencing technologies, we have analyzed the exome sequences of brain tumor patients with matched tumor and normal blood. We show that mutations contained in sequencing data can be revealed using this technique thus providing a new methodology for determining subsequences of given length containing mutations i.e. its value differs from those of subsequences without mutations. A potential application for this technique involves simplifying the procedure of finding segments with mutations, speeding up genomic research, and accelerating its implementation in clinical diagnostic. Moreover, the prediction of a mutation associated to a family of frequent mutations in numerous types of cancers based purely on the value of the Kolmogorov function, indicates that this applied marker may recognize genomic sequences that are in extremely low abundance and can be used in revealing new types of mutations.  

\vspace{0.1in}


\newpage 

\section{Introduction}

To study mutations in the genomic sequences of cancerous tissues we use the statistic introduced initially by Kolmogorov \art{K} and later developed by Arnold \art{Arn1},\art{Arn2} in defining a degree of randomness (stochasticity) for a given sequence of real numbers. The universality of the method has been revealed at measuring the degree of randomness of finite sequences in theory of dynamical systems and in number theory \art{Arn1}. This approach has been applied to physical problems, i.e. at the study of non-Gaussianities in cosmic microwave background radiation \art{GK},\art{Kmap},\art{Planck}, of X-ray galaxy clusters \art{Xray}. This method was instrumental in detecting the thermal trust effect (Yarkovsky-Rubincam effect) for the first time in the properties of satellites probing General Relativity \art{Lag}.

\section{Method}

Let us briefly introduce the technique and the descriptors which were then applied to the genomic data. For $\{X_1,X_2,\dots,X_n\}$  $n$ independent real-valued variable ordered in increasing manner $X_1\le X_2\le\dots\le X_n$ the {\it cumulative distribution function} (CDF) is defined as $F(x) = P\{X\le x\}$ \art{K},\art{Arn1},\art{Arn2}.

The {\it empirical distribution function} $F_n(x)$ will be

\bea
F_n(x)=
\begin{cases}
0\ , & x<X_1\ ;\\
k/n\ , & X_k\le x<X_{k+1},\ \ k=1,2,\dots,n-1\ ;\\
1\ , & X_n\le x\ .
\end{cases}
\eea

Then the stochasticity parameter is defined as
\be
\lambda_n=\sqrt{n}\ \sup_x|F_n(x)-F(x)|\ .
\ee
Kolmogorov's theorem \art{K} states that for any continuous {\bf CDF} $F$ the following limit is converged uniformly
\be
\lim_{n\to\infty}P\{\lambda_n\le\lambda\}=\Phi(\lambda)\ ,
\ee
where the $\Phi(0)=0$,
\be
\Phi(\lambda)=\Sum_{k=-\infty}^{+\infty}\ (-1)^k\ e^{-2k^2\lambda^2}\ ,\ \  \lambda>0\ ,\label{Phi}
\ee
and the distribution (Kolmogorov's) $\Phi$ is independent on $F$. For small values of $\lambda$ the following approximation yields 
\be
\Phi(\lambda) \approx \frac{\sqrt{2\pi}}{\lambda} e^{-\frac{{\pi}^2}{8\lambda^2}}.
\ee

This method thus provides the measure of the degree of randomness (stochasticity) for sequences of $n$ values within the interval of $\lambda_n$ [0.3, 2.4] \art{Arn1},\art{Arn2}, see also \art{atto}.

\section{Data}

The following data have been used for the analysis. Gliomas are the most frequent malignant tumors of the CNS and are defined by WHO grade I to grade IV classification standards and histopathological features \art{Dolecek},\art{Louis}. 
Application of novel techniques to elucidate the fundamental genetic mutations in Grade II-III astroctyomas and Grade IV glioblastomas is a critical next step in glioma research. Here, we interrogated exome data of 30 brain tumor patients from the Preston Robert Tisch Brain Tumor Center at Duke University as described previously  \art{Duke1}. The exomes of 30 patients were selected as they provided a large enough dataset to conduct our initial analysis.  Each case contained four datasets corresponding to aligned paired end sequencing data files (both a forward and reverse file for each patients tumor and normal blood). 
Included in this study were 15 grade III astrocytomas and 15 grade IV glioblastoma. Samples on average yielded 36 and 32 somatic mutations for grade III astrocytomas and grade IV glioblastoma, respectively. Of particular interest to the general cancer community is the prevalence of highly recurrent mutations across all types of cancer. To this end, a list of highly recurrent mutations occurring in 23 genes commonly seen in cancer was used to interrogate the dataset (Table 1). The selected genes are from a list of frequently mutated genes in cancer provided by Personal Genome Diagnostics (Baltimore, MD). Next, we surveyed the exome data to identify any occurrence of these 407 highly recurrent mutations located within these 23 commonly mutated genes, see  \art{Jones}.

\begin{center}
\begin{table}
{
\footnotesize
\begin{tabular}{lll}
\hline
Gene   & Gene                                              &Transcript     \\
Symbol & Description                                       &Accession      \\
\hline
\hline
ABL1   & c-abl oncogene 1; non-receptor tyrosine kinase    &X16416         \\
AKT1   & v-akt murine thymoma viral oncogene homolog 1     &ENST00000349310\\
AKT2   & v-akt murine thymoma viral oncogene homolog 2     &NM 001626.3    \\
ALK	   & anaplastic lymphoma receptor tyrosine kinase      &NM 004304      \\
BRAF   & v-raf murine sarcoma viral oncogene homolog B1    &NM 004333      \\
CDK4   & cyclin-dependent kinase 4                         &NM 000075.2    \\
EGFR   & epidermal growth factor receptor                  &NM 005228      \\
ERBB2  & v-erb-b2 erythroblastic leukemia viral            &               \\
       & oncogene homolog 2                                &NM 004448      \\
FGFR1  & fibroblast growth factor receptor 1	             &NM 023110      \\
FGFR3  & fibroblast growth factor receptor 3               &NM 000142      \\
FLT3   & fms-related tyrosine kinase 3                     &NM 004119      \\
HRAS   & v-Ha-ras Harvey rat sarcoma viral                 &               \\
       & oncogene homolog                                  &NM 005343      \\
IDH1   & isocitrate dehydrogenase 1 (NADP+); soluble       &NM 005896.2    \\
IDH2   & isocitrate dehydrogenase 2 (NADP+); mitochondrial &NM 002168.2    \\
JAK2   & Janus kinase 2                                    &ENST00000381652\\
KIT	   & v-kit Hardy-Zuckerman 4 feline sarcoma viral      &               \\
  	   & oncogene homolog                                  &NM 000222      \\
KRAS   & v-Ki-ras2 Kirsten rat sarcoma viral               &               \\
       & oncogene homolog                                  &NM 004985      \\
MET	   & met proto-oncogene (hepatocyte growth )           &               \\
  	   & factor receptor)                                  &NM 000245      \\
MET	   & met proto-oncogene (hepatocyte growth             &               \\
  	   & factor receptor)                                  &NM 001127500   \\
NRAS   & neuroblastoma RAS viral (v-ras) oncogene homolog  &NM 002524      \\
PDGFRa & platelet-derived growth factor receptor;          &               \\
       & alpha polypeptide                                 &NM 006206      \\
PIK3CA & phosphoinositide-3-kinase; catalytic;             &               \\
       & alpha polypeptide                                 &NM 006218.1    \\
RET	   & ret proto-oncogene                                &NM 020975      \\
\hline
\end{tabular}
\caption{23 genes commonly mutated in cancer.}
}
\end{table}
\end{center}

{

\section{Analysis}

The data for 30 patients have been analysed in the following manner. First, for each dataset the cumulative distribution function has been obtained for 3-combinations of nucleotides (guanine, adenine, thymine, and cytosine (G,A,T,C) i.e. of codons.
For comparison, in one and two-based analysis of the same data, no CDF was possible to define due to large scatter in the frequency counts, however it was possible for the 3-base empirical distribution for each particular genomic sequence, followed by obtaining of the stochasticity parameter and then the Kolmogorov's function $\Phi$. The fact that $\Phi$ was possible to define only for 3-base CDF can be considered as a genuine feature linked to the nature of the genetic coding (first noticed by cosmologist George Gamow in 1950s) determined by the chemical properties of the molecules forming the nucleotides.

The above described exome data were represented in a format of over 50 mln rows, of 100-base each, i.e. total in about $10^9$-valued sequences. We analysed such datasets for 30 patients, a block of 4 datasets, representing paired-end sequencing,  was available for each patient, including those for blood and tumor, denoted as normal and with tumor, respectively. The following aim was inquired into: whether the K-A technique is able to distinguish the strings i.e. sequence pieces of given length, with and without mutations, for a given sample of mutations. Figure 1 represents the results for computations for 100-base rows containing such mutations in all 119 blocks (dark column in the right; 1 file was corrupted), and the mean of the function $\Phi$ for 50 rows without mutations (light-colored column in the left) in the same sequence where the former mutations have been located. The same procedure has been repeated for shorter i.e. for 50 and 25-base strings with and without mutations (the right two double-columns, respectively). The shorter, 50-base strings were generated in the following manner: if the mutation is located completely either in the first or the second half of the 100-base string, then the corresponding halves were included in the analysis. In the case of partial location, the proper number of bases was included from either side, the mutations are included completely even at partly passage to the next row. Similar was the case for 25-base strings, while for no-mutation strings (rows) their initial 50 or 25-base parts were included in the analysis. For no-mutation rows their alignment i.e. their position in the sequence was not important. The error bars correspond to standard errors. The analysis was performed by means of a software created in {\it Pascal} in environment {\it Delphi} and intended to be made public in due course. The CPU time for one sequence (about $10^9$ nucleotides) is about 1 hour for i7, 2600 3.4 GHz processor of 6GB memory.

\begin{figure}[ht]
\begin{center}
\includegraphics[width=\textwidth]{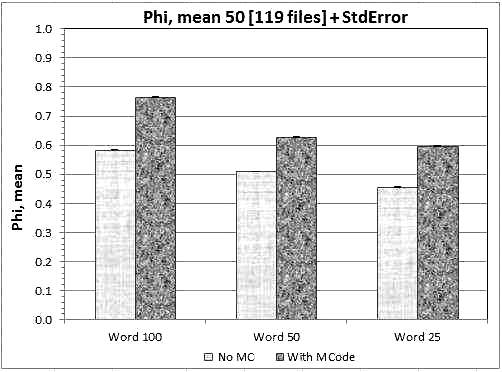}
\caption{The function $\Phi$ averaged for rows with mutations (dark bars) and normal ones (light bars) averaged over 50 rows, for 100, 50 and 25-base rows (here denoted as word), correspondingly. Error bars correspond to standard errors.}
\label{phi1}
\end{center}
\end{figure}

\begin{table}
{
\caption{The codes (MC) and frequency counts for 7 highest recurrent mutations in the studied database.}
\footnotesize
\begin{center}
\renewcommand{\baselinestretch}{1.2}
\renewcommand{\tabcolsep}{2.5mm}
\ttfamily
\begin{tabular}{ccc}
\hline
 			N    &     MC   &   Frequency \\		
\hline
\hline
1   &    AAAGANAAATT  &   3032 \\
2   &    CATCTNAAAAA  &   2569 \\
3   &    AGAAGNAAAAA  &   2427 \\
4   &    ATTTTNTCTTT  &   2409 \\
5   &    TGCCCNGGCTG  &   2253 \\
6   &    CTTTCNTTTCT  &   2200 \\
7   &    TCTTCNTTTTT  &   2123 \\
\hline
\end{tabular}
\end{center}
}
\end{table}

\begin{figure}[ht]
\begin{center}
\includegraphics[width=\textwidth]{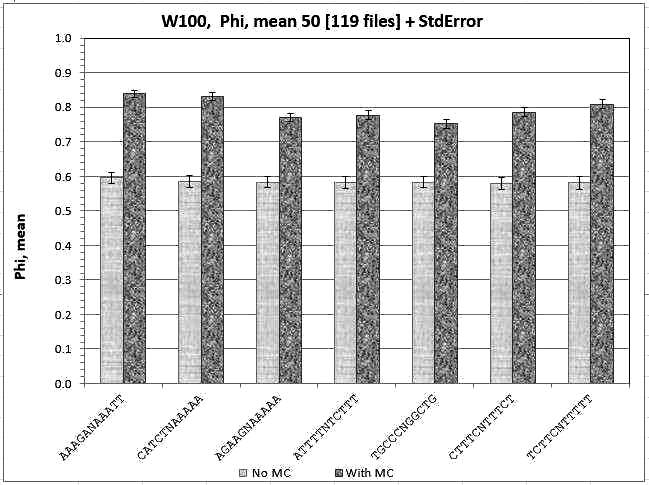}
\caption{The same as in Figure 1, but with the averaged values of the function $\Phi$ for 100-base rows for the highest recurrent specific mutations in the studied dataset as listed in Table 2.}
\label{phi2}
\end{center}
\end{figure}

Then, rows containing the most frequent specific mutations in the same dataset of 30 patients (Table 2), have been analysed. Obviously, more frequent mutations provide higher statistics, and Table 2 and Figure 2 are represented to show the scales of the input frequency numbers vs the results. The mutations i.e. the genes and the mutant positions and amino acid changes are known for the codes listed in Table 2, and from the individual mutation reports of the performed studies \art{Duke1} one can list all mutations contained within each tumor; we intend to address these issues in next publications on the applications of method discussed here. The results for the mean $\Phi$ with standard error bars are presented  in Figure 2. One can see that certain specific mutations can be distinguished by the value of mean $\Phi$.

\section{One step further: possibility for discovering new mutations}

Up to now we were estimating the Kolmogorov function $\Phi$ for genomic sequence pieces (rows) with known mutations and without (of those) mutations. We reveal the differences in the mean values of $\Phi$ for rows with and without those mutations. If so, then one can pose the inverse problem, namely, can one try to detect unknown mutations based on the value of $\Phi$, if estimated blindly in a given genome sequence. In Table 4 we show the results for a sample from the above studied database from genome with tumor (081T1): 11 rows (of over 50 mln) with $\Phi > 0.7$ (only the rows with the number of unidentified nucleodites $N \leq 3$ have been taken into account), have been revealed with the codes given in the Table 3 as candidates for mutations. Obviously, certain part of these cases can be just noise, i.e. without any association to real mutations, however, if at least part of these list when studied by conventional methods will be confirmed as associated to mutations, then one will have an explicit tool for detection of unknown mutations by means of this relatively simple, i.e. of small time and manpower consuming method. Certainly, the exhausting answer to this question will need comprehensive parallel studies with different analysis methods. However, for now, for illustration, for the given sample of detected candidates for mutations, we performed the following. We have sampled five reads from the aforementioned list of 11 rows in Table 3 and aligned them to the human genome for further investigation. While none of these five reads matched to previously reported mutations in the COSMIC database, one of the reads aligned to a region of interest for oncogenomic laboratories. This read aligned to the transcriptional regulator ARID1A, a SWI/SNF family member that is frequently mutated in numerous types of cancers, including gastric, ovarian, and pancreatic cancers \art{Jones1},\art{Sh},\art{Wa},\art{Wi}. Probability for a chance coincidence is less than $10^{-5}$ (assuming for simplicity an equipartition in the frequency counts, cf. Table 2), thus proving the efficiency of the method for blind application to detasets using the prviously calibrated $\Phi$.

Another potential application for this technique involves detection of rare variants of sequencing data, where the parameter $\Phi$ may recognize genomic sequences that are in extremely low abundance and warrant further investigation by the researchers.

\begin{center}
\begin{table}
{
\caption{The codes for candidates for mutations in a genome sequence with high $\Phi$ at given row numbers.}
\renewcommand{\baselinestretch}{1.0}
\renewcommand{\tabcolsep}{5.5mm}
\small
\ttfamily
\begin{tabular}{rc}
\hline
    Line N &         MC       \\		
\hline
\hline
   1186009 &      TTGTGNAAGGG \\
   4073568 &      CCACGNCCTGG \\
  11648505 &      ATACANAACGC \\
  21240249 &      AAGGANACTGA \\
  21827969 &      CAATTNGGGAA \\
  33372865 &      GCCGGNCGCGG \\
  34549019 &      TGGCCNAGAAG \\
  36995074 &      TGAAGNGTTCT \\
  42622891 &      TTGTTNTTTTA \\
  43978737 &      AGAAANATATT \\
  50647272 &      TGACTNAAAGG \\
\hline
\end{tabular}
}
\end{table}
\end{center}

\section{Conclusions}

The following basic conclusions can be drawn from the above analysis:

(a) the stability of the descriptor, that is small standard errors and hence high and stable confidence level of the values of $\Phi$ for paired-end sequence rows both for normal, i.e. without mutations, and those with mutations;

(b) the difference in values of $\Phi$ for rows with and without mutations;

(c) the considered variations of string lengths still reflect the tendency;

(d) rows with certain mutations can be distinguished by means of the used marker.

The presented results demonstrate for the greater cancer research community the power of the Kolmogorov-Arnold technique for identification of mutations in paired-end genome sequencing data. In addition to the significance of revealing the important nature of the difference in the degree of the randomness between the genome sequence with and without mutations, the method also has an important application potential. It may be applied to non aligned sequencing segments, which may significantly simplify procedure of finding segments with mutations and could speed up genomic research and its implementation in clinical diagnostic.

Finally, the consideration of the inverse problem, namely, the revealing of a mutation associated to a tumour based purely on the computation of the value of the marker when blinded to the data, indicates that the latter may be used for detection of rare or new types of mutations. 

\vspace{0.2in}

• Competing interests. Authors have no competing interests.

• Authors' contributions. VGG, AK, SS, GY carried out the statistical analysis, drafted the manuscript; PK, GV, ZR, GM collected field data; VGG, HY, GV, PK, BV conceived of the study, coordinated the study and helped draft the manuscript. All authors gave final approval for publication.

• Acknowledgments. We acknowledge the fruitful discussions with Y. He and A. Tovmasian.

• Funding statement. VGG, AK, SS, GY, GM, BV are funded by NSF (HRD-0833184) and NASA (NNX09AV07A).  HY, GV, PK, GV, ZR are funded by The V Foundation and an Accelerate Brain Cancer Cure Foundation grant, a Voices Against Brain Cancer Foundation grant, a Pediatric Brain Tumor Foundation Institute grant, a James S. McDonnell Foundation grant, American Cancer Society Research Scholar Award RSG-10-126-01-CCE, and National Cancer Institute Grant 5R01-CA140316.

\end{document}